\documentclass{article}
\usepackage[utf8]{inputenc}
\usepackage[margin=1in]{geometry}
\usepackage{amsmath, amsfonts}
\usepackage{apacite}
\usepackage{natbib}
\usepackage{hyperref}
\hypersetup{colorlinks=false,breaklinks=true,hidelinks}
\usepackage{cleveref}
\usepackage{graphicx}
\graphicspath{{./figures/}}
\DeclareGraphicsExtensions{.pdf}
\usepackage{authblk}
\usepackage{float}
\usepackage{subcaption}
\usepackage{booktabs, multirow}
\usepackage{url}
\usepackage{nicefrac}
\usepackage[table,xcdraw]{xcolor}
\newcolumntype{L}[1]{>{\raggedright\let\newline\\\arraybackslash\hspace{0pt}}m{#1}}
\newcolumntype{C}[1]{>{\centering\let\newline\\\arraybackslash\hspace{0pt}}m{#1}}
\newcolumntype{R}[1]{>{\raggedleft\let\newline\\\arraybackslash\hspace{0pt}}m{#1}}

\title{\textbf{An End-to-End Earthquake Detection Method for Joint Phase Picking and Association using Deep Learning}}

\author[1]{Weiqiang Zhu\textsuperscript{*}}
\author[2]{Kai Sheng Tai\textsuperscript{*}}
\author[1]{S. Mostafa Mousavi}
\author[2]{Peter Bailis}
\author[1]{Gregory C. Beroza}
\affil[1]{\small Department of Geophysics, Stanford University, Stanford, CA, 94305}
\affil[2]{\small Computer Science Department, Stanford University, Stanford, CA, 94305}
\date{}

\begin{document}

\maketitle

\begin{abstract}
Earthquake monitoring by seismic networks typically involves a workflow consisting of phase detection/picking, association, and location tasks. In recent years, the accuracy of these individual stages has been improved through the use of machine learning techniques. In this study, we introduce a new, end-to-end approach that improves overall earthquake detection accuracy by jointly optimizing each stage of the detection pipeline. We propose a neural network architecture for the task of multi-station processing of seismic waveforms recorded over a seismic network. This end-to-end architecture consists of three sub-networks: a backbone network that extracts features from raw waveforms, a phase picking sub-network that picks P- and S-wave arrivals based on these features, and an event detection sub-network that aggregates the features from multiple stations and detects earthquakes. We use these sub-networks in conjunction with a shift-and-stack module based on back-projection that introduces kinematic constraints on arrival times, allowing the model to generalize to different velocity models and to variable station geometry in seismic networks. We evaluate our proposed method on the STanford EArthquake Dataset (STEAD) and on the 2019 Ridgecrest, CA earthquake sequence. The results demonstrate that our end-to-end approach can effectively pick P- and S-wave arrivals and achieve earthquake detection accuracy rivaling that of other state-of-the-art approaches.
\end{abstract}

\section{Introduction}

\begin{figure}
    \centering
    \includegraphics[width=\textwidth]{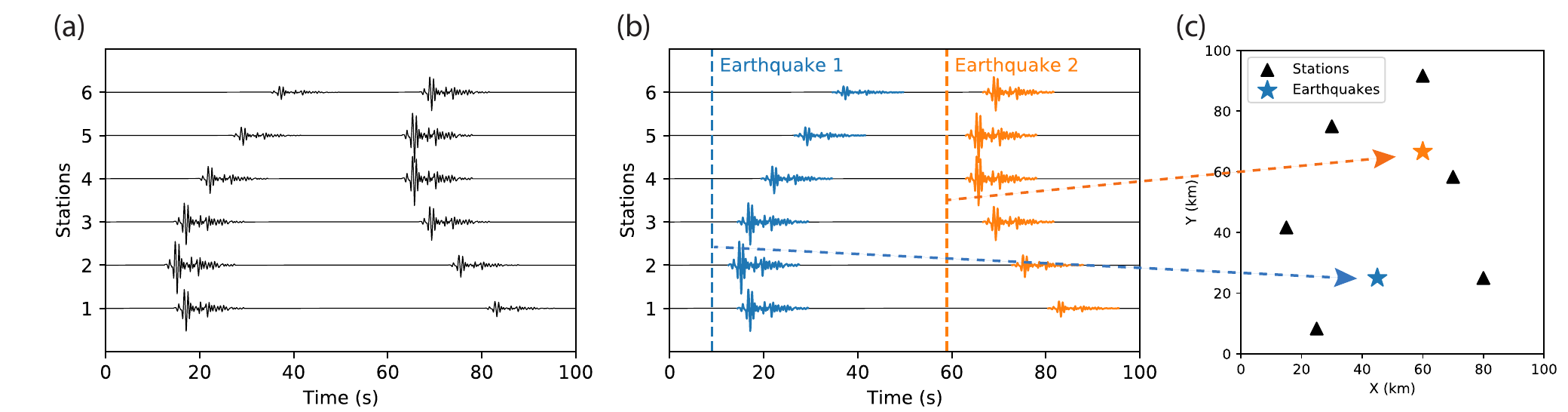}
    \caption{Schematic of earthquake event detection over a seismic network: (a) continuously recorded seismic data; (b) multi-stage tasks including: detecting single-station signals, associating signals from a common earthquake, and (c) estimating earthquake source parameters, such as origin time and location.}
    \label{fig:illustration}
\end{figure}

Earthquakes are routinely monitored by local and global seismic networks, which consist of several to hundreds of seismographs that continuously record ground motion. Earthquake monitoring agencies process the recorded seismic waveforms to detect and catalog earthquakes. Most earthquake monitoring workflows involve serial processing through a sequence of stages, which includes seismic phase detection/picking, association, and event location. First, a phase detection/picking algorithm identifies P- and S-phases independently at each station. An association algorithm then aggregates these phases from several stations by determining whether the phase arrival times are consistent with travel-times from a common earthquake hypocenter. The associated phases are then used to determine earthquake location, magnitude, and other properties related to the earthquake source~(\Cref{fig:illustration}).
To improve earthquake detection accuracy, especially for frequent small earthquakes, previous research has focused on independently optimizing each individual task, for example by increasing phase detection/picking sensitivity \citep{ross2018generalized, zhu2019phasenet} and improving phase association robustness \citep{yeck2019glass3, zhang2019rapid}.

Traditional algorithms for phase detection/picking from continuous seismic waveforms use human-selected characteristic features (e.g., changes of amplitude, frequency, and other statistical properties of the time series) to detect the presence of a seismic signal within background noise and to determine accurate arrival times~\citep{allen1978automatic, bai2000automatic, saragiotis2002pai, lomax2012automatic, baillard2014automatic, ross2014automatic, mousavi2016automatic, mousavi2016fast}. Recently, deep learning has emerged as an effective method to learn feature representations of seismic signals automatically by training on large historical datasets~\citep{perol2018convolutional, ross2018generalized, ross2018p, zhu2019phasenet, mousavi2019cred, zhou2019hybrid, zhu2019deep, mousavi2020earthquake}. In particular, deep-learning-based phase detectors and pickers have been shown to significantly improve the detection rate for small earthquakes~\citep{park2020machine, ross20203d, wang2020injection, tan2021machine, beroza2021machine}. However, most deep-learning-based phase detectors and pickers use only single-station information and ignore the contextual information provided by other stations. Considering multiple stations thus becomes a potential direction to improve phase detection performance.

Once phase detections are produced at each station, they are associated across multiple stations in a seismic network. Association methods aggregate these single-station detections into a set of seismic events and filter out false positives in the process. Most association methods are based on the idea of back-projection, which aggregates phase detections that fit a theoretical travel-time moveout given a hypothetical event source and a wavespeed model~\citep{dietz2002notes, johnson1997robust, patton2016hydra,  zhang2019rapid, yeck2019glass3}. An event detection is declared when a number of phases are associated to a common source event. Several deep-learning-based methods have been proposed to learn phase arrival time moveout patterns~\citep{ross2019phaselink} or waveform similarity~\citep{mcbrearty2019pairwise, dickey2020beyond} to improve association. 

Treating phase detection and association independently can limit the overall accuracy of the event detection pipeline. First, accurate phase detection/picking at a single station is a difficult task for the low signal-to-noise ratio (SNR) arrivals of small-magnitude events, whose numbers dominate earthquake catalogs. Second, association using only picked phase times does not exploit potentially informative waveform features across stations. Since weak phases that fall below the detection threshold are filtered out in the first step, the information they carry is lost to subsequent processing and thus cannot contribute to the association step for small events.

An alternative approach to this multi-stage earthquake monitoring workflow of single-station phase detection and multi-station phase association is to develop array-based event detection methods, which can improve the sensitivity for events that are too weak to be detected reliably by a single station. Methods like template matching~\citep{gibbons2006detection, shelly2007non, zhang2015effective} and shift-and-stack~(i.e., back-projection or beamforming) \citep{kao2004source, kiser2013hidden, li2018high} exploit the coherent waveform signals of multiple stations to enhance detection sensitivity. However, these array-based methods have several disadvantages. Template matching requires \textit{a priori} information in the form of a catalog of detected events as templates and if done comprehensively, can have a high computational cost \citep{ross2019searching}. Back-projection relies on high similarity between waveforms. Sequences of tiny earthquakes usually have complex waveforms contaminated by noise, which when coupled with subsurface heterogeneity, complex wave propagation, and wave-speed model uncertainty, lead to a smeared back-projection image from which it can be challenging to extract events. 
Several deep-learning-based methods have treated an array of seismic recordings as an image and applied 2D convolution neural networks to pick phases or detect events \citep{zheng2020sc, yang2021simultaneous, shen2021array, zhang2021generalized}. This approach is effective for a fixed geoemtry of seismic recordings, such as a shot gather in exploration seismology, but it unsuitable for the important problem of long-term seismic monitoring for which station geometry will vary with time. This common situation would lead to poor model generalization when applied to a seismic network geometry that differs substantially from that of the seismic network(s) used during training. 

We present a novel approach designed to combine the advantages of effective representation learning by deep neural networks and robustness of array-based event detection. This method, which we call EQNet, combines feature extraction, phase picking, and event detection stages in an end-to-end neural network architecture in order to incorporate knowledge of the downstream association task into the detection step and help the network to learn features that are effective for multiple tasks. The architecture first extracts features from seismic waveforms recorded at each seismic station using a deep backbone network. These feature vectors are then processed by two sub-networks for picking P- and S-phases and for aggregating features from multiple stations and detecting earthquake events. This multi-task approach was also used by \citet{mousavi2020earthquake} for combining phase picking and event detection tasks at a single station. A characteristic of our approach is that we \textit{jointly} optimize the neural network parameters of feature extraction, phase picking, and event detection to maximize the earthquake detection performance using the entire seismic network. This avoids the need for hand-designed characteristic functions and association rules used in conventional multi-stage approaches and preserves information in the input waveforms through all stages. We benchmarked the phase picking performance on the STanford EArthquake Dataset (STEAD) \citep{mousavi2019stanford} and evaluated the earthquake detection performance against four state-of-the-art catalogs previously developed for the 2019 Ridgecrest, CA earthquake sequence. Our end-to-end approach provides a promising new direction for further improving earthquake monitoring using deep learning.

\section{Method}

\subsection{End-to-end architecture}

\begin{figure}
    \centering
    \includegraphics[width=\textwidth]{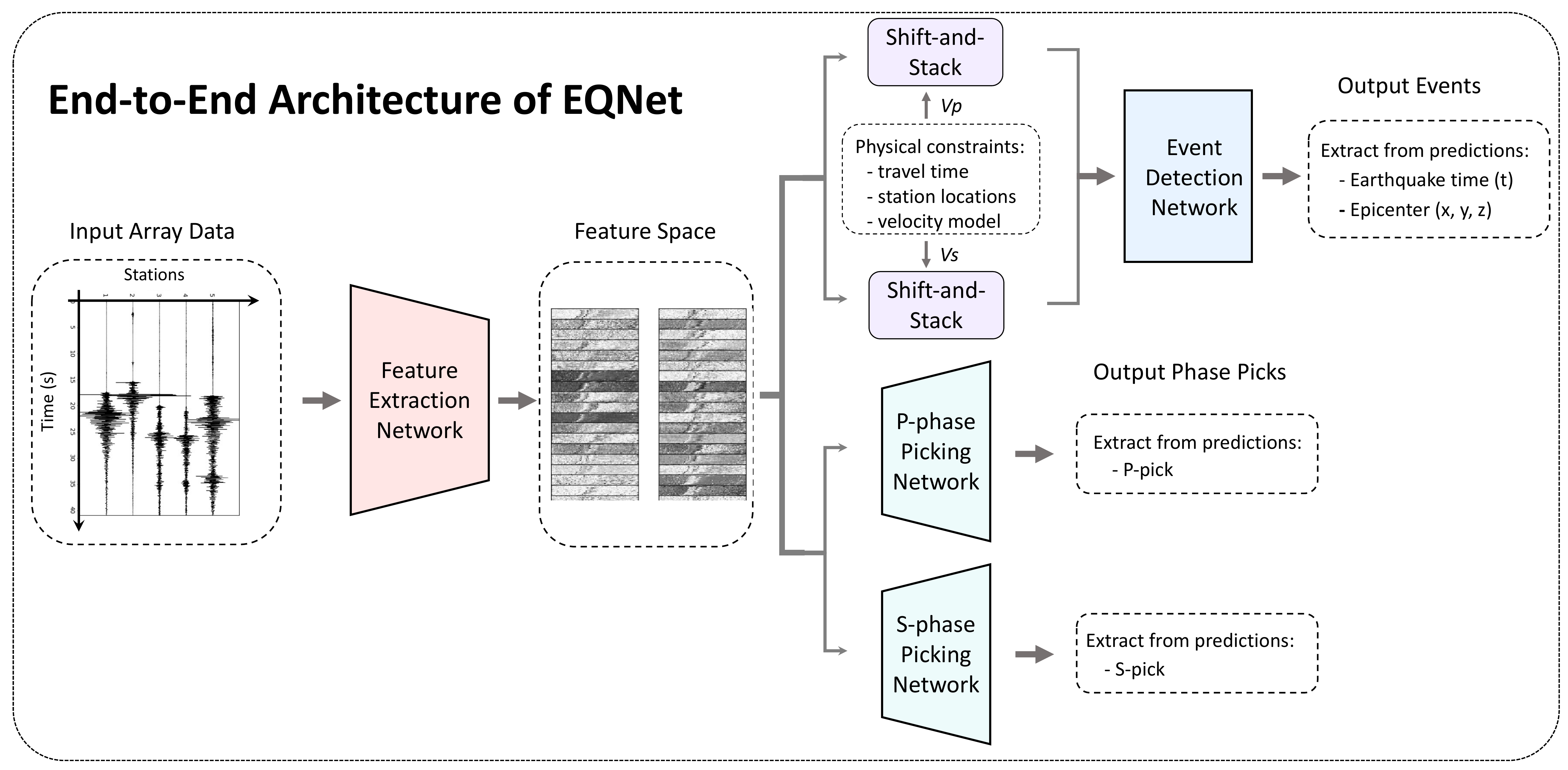}
    \caption{Architecture of the end-to-end earthquake detection model (EQNet). The input data are seismic waveforms recorded by multiple stations in a seismic network.  The outputs are two activation sequences of P- and S-picks and an activation map of earthquake events with approximate earthquake time and location. EQNet consists of four sub-modules: feature extraction, phase picking, shift-and-stack, and event detection. The feature extraction network transforms raw seismic waveforms into feature representations using a 1D ResNet-18 model. The phase picking network then predicts two activation sequences for P- and S-phase arrivals based on the features. The shift-and-stack module is designed to sample candidate earthquake locations, calculate travel-times at each station location, and shift the features accordingly, allowing generalization to different station locations and seismic wavespeed models. The event detection network predicts an activation map for approximate earthquake times and locations based on the shifted features. These three networks are optimized simultaneously during training to improve earthquake detection performance. }
    \label{fig:end2end-model}
\end{figure}

\begin{figure}
    \centering
    \includegraphics[width=\textwidth]{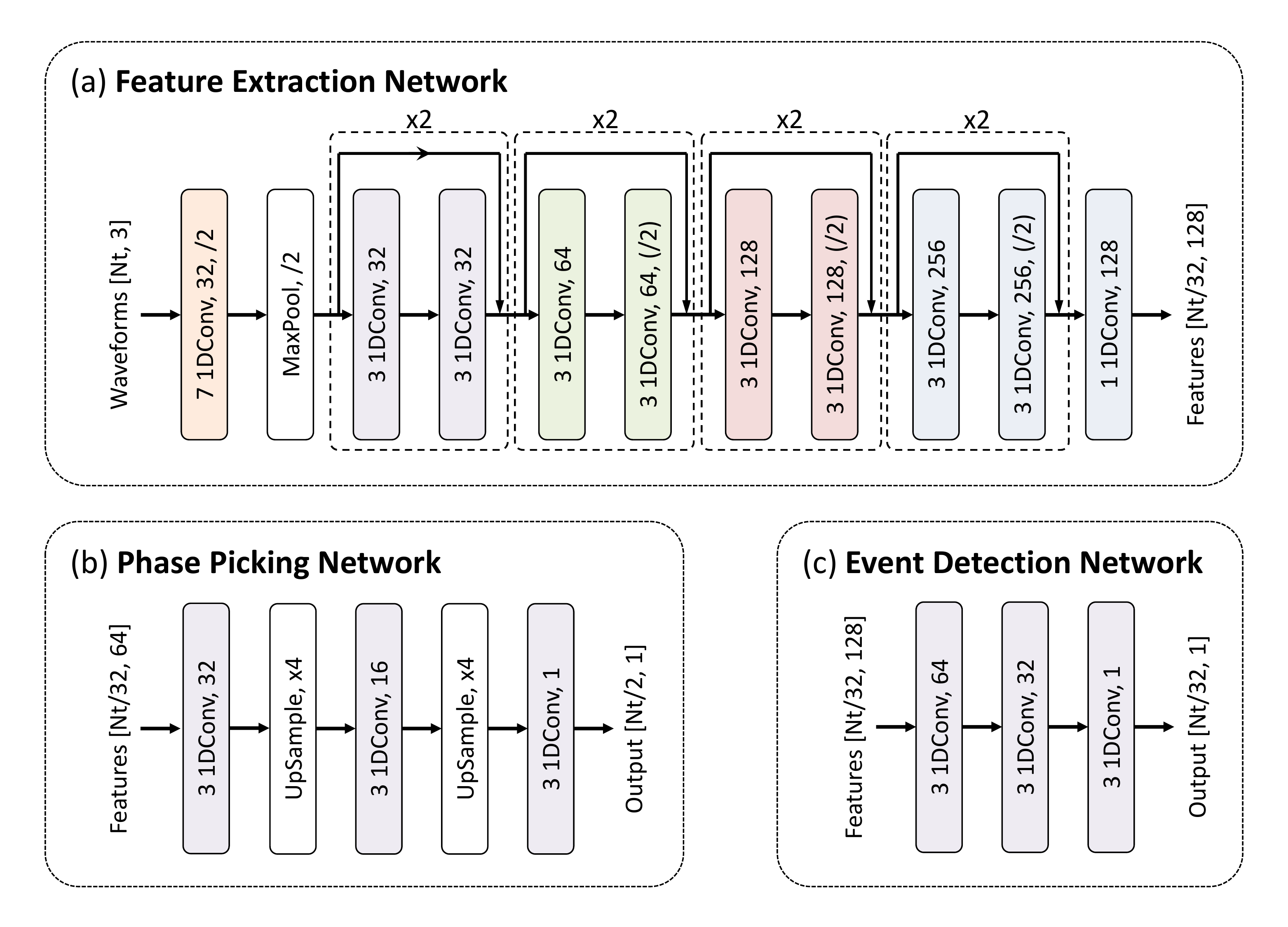}
    \caption{Neural network parameters: (a) a backbone neural network for feature extraction; (b) a sub-network for phase picking; (c) a sub-network for event detection. Numbers in the square brackets are the shapes of input and output data. For example, the input waveform has a length of $Nt$ samples from 3 components. The numbers inside each neural network layer are the convolution kernel size, the number of channels, and the stride step (inside brackets). Batch normalization layers and ReLU activation functions are used after each 1D convolutional layer but are not plotted here.}
    \label{fig:network-parameter}
\end{figure}

To perform the phase picking, association, and event detection tasks in an end-to-end fashion, we design a novel architecture for EQNet (\Cref{fig:end2end-model}) that consists of three sub-networks, including a backbone feature extraction network, a phase picking network, and an event detection network, as well as a shift-and-stack module.
EQNet processes a collection of seismograms from multiple stations as input, and produces both picked P/S-phases and detected earthquake events with rough estimates of earthquake origin time and location as outputs. 
The backbone feature extraction network maps raw seismic waveforms into a feature space with a condensed time dimension. The feature extraction network is modified from the 18-layer residual network (ResNet-18) \citep{he2016deep} that uses 1D convolution to process three-component seismic waveforms and automatically extract condensed features. 
The extracted features are then processed by two sub-networks. The phase picking network extracts P- and S-phase picks from these features, and the event detection network detects earthquake events from shifted features produced by the shift-and-stack module. The shift-and-stack module performs a transformation in the feature domain, similar to the back-projection process, by sampling candidate hypocenters, shifts the features based on estimated theoretical travel-times, and generates a collection of aligned features. The module introduces prior knowledge of physical constraints on arrival time moveout, epicentral distance, and the wavespeed model. The event detection network then classifies whether an earthquake exists at a specific candidate location and time based on the aligned views of features. The feature extractor network, phase picker and event detector sub-networks serve functions similar to the backbone network and multiple head networks in object detection methods, such as YOLO \citep{bochkovskiy2020yolov4}. During training, we jointly optimize the parameters of these three networks using a summation of three binary cross-entropy losses as a multi-task optimization target:
\begin{align}
    \mathcal{L} &= \lambda_P \mathcal{L}_P + \lambda_S \mathcal{L}_S + \lambda_{EQ} \mathcal{L}_{EQ} \\
    \mathcal{L}_P &= -\sum_{t=1}^T y_{P,t}\log \hat{y}_{P,t} + (1-y_{P,t})\log (1 -  \hat{y}_{P,t}) \\
    \mathcal{L}_S &= -\sum_{t=1}^T y_{S,t}\log \hat{y}_{S,t} + (1-y_{S,t})\log (1 -  \hat{y}_{S,t})  \\
    \mathcal{L}_{EQ} &= -\sum_{t=1}^T y_{EQ,t}\log \hat{y}_{EQ,t} + (1-y_{EQ,t})\log (1 -  \hat{y}_{EQ,t})
\end{align}
where $\mathcal{L}_P$ and $\mathcal{L}_S$ are the losses of the phase picking networks for P- and S-phase respectively, and $\mathcal{L}_{EQ}$ is the loss of the event detection network. $T$ is the number of time points, $y$ is the ground truth label, and $\hat{y}$ is the network prediction. $(\lambda_P, \lambda_S, \lambda_{EQ})$ are weights for each loss function. For this proof-of-concept study, we do not tune for the optimal weighting, but set all weights to 1.
The parameters of these three neural networks are shown in \Cref{fig:network-parameter}. 
We use two phase picking sub-networks for picking P- and S-phases, and each network takes half of the extracted features as input. These two sets of features are shifted using P- and S-wave velocities respectively and processed by the event detection sub-network.
We train this end-to-end neural network model using the AdamW optimizer \citep{loshchilov2017decoupled}, a weight decay rate of $1\times10^{-4}$, and a cosine learning rate decay strategy \citep{loshchilov2016sgdr} with an initial learning rate of $3\times10^{-4}$.

\subsection{Dataset and training}

\begin{figure}
    \centering
    \includegraphics[width=\textwidth]{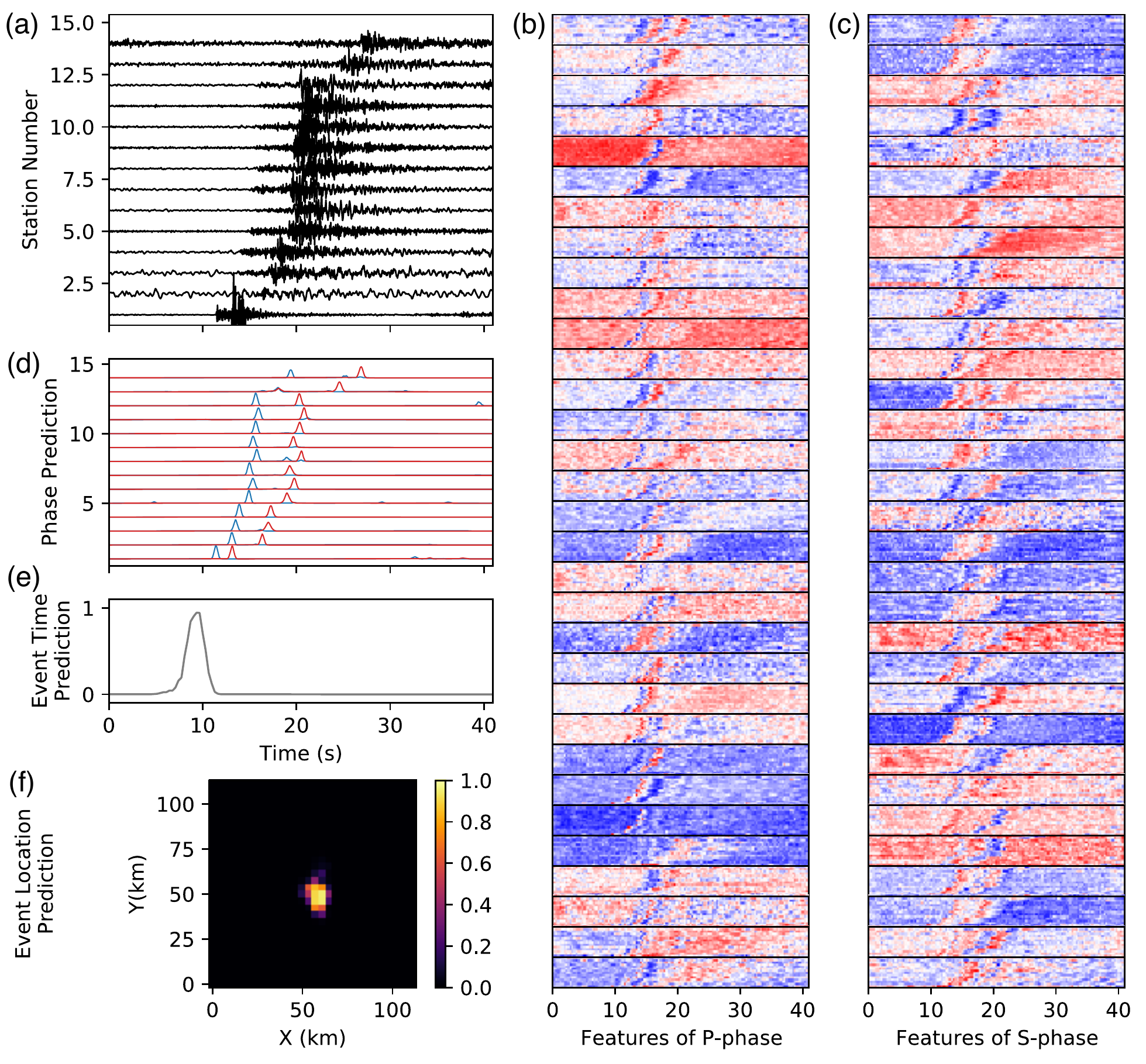}
    \caption{Prediction example: (a) seismic waveforms; (b, c) features of P-phase and S-phase extracted by the backbone feature extraction neural network. Each sub-panel shows one output channel of the backbone network. (d) P- and S-phase activation scores predicted by the phase picking neural network. (e, f) Earthquake activation scores predicted by the event detection neural network. Specifically, (e) shows the max score along the time axis from which we extract the event time, and (f) shows the max score along spatial axes from which we extract the event location.}
    \label{fig:prediction-sample}
\end{figure}

We collected a training set from the Northern California Earthquake Data Center \citep{ncedc2014northern} and selected events with manual P- and S-phase picks from more than 3 stations. The total training dataset contains 99,465 events and 563,790 P- and S-picks. \Cref{fig:prediction-sample}(a) shows one example recorded at 14 stations. We use the same truncated Gaussian-shaped target function used for PhaseNet \citep{zhu2019phasenet} as training labels for P/S-phase arrival times and earthquake origin time. The Gaussian-shape target function allows some uncertainty in manual labels and balances the distribution between a small number of positive sample points of manual labels and the remaining majority negative sample points in a seismic waveform to improve training accuracy and speed. We set two different widths of the Gaussian-shaped target functions: 1s for phase arrival time and 2s for earthquake origin time to account for different uncertainty levels. \Cref{fig:prediction-sample}(d) and (e) show the corresponding prediction scores after training of the phase picking network and the event detection network respectively. We use a negative sampling approach \citep{goldberg2014word2vec} to collect positive samples from the correct earthquake location and negative samples from other random locations. This negative sampling process helps balance the sparse space of true earthquake locations relative to the much larger space of candidate locations spanned by the entire monitoring area to speed up the training. During inference, we take the continuous seismic waveforms ($Nt$) as input data. The shift-and-stack module uniformly but coarsely samples the whole space at $Nx \times Ny$ grid points with a horizontal interval of $\sim$4 km. \Cref{fig:prediction-sample}(f) shows the corresponding activation map of prediction scores. From the spatial-temporal predictions above a threshold, we first extract earthquake times from peaks along the time axis (\Cref{fig:prediction-sample}(e)) and then determine the earthquake location using the geometric median of the top 20 activated grids (\Cref{fig:prediction-sample}(f)). We visualize the automatically extracted features in \Cref{fig:prediction-sample}(b) and (c). Each mini-panel shows one output channel of the feature extraction network in \Cref{fig:network-parameter}(a). The horizontal axis presents time, and the vertical axis represents stations. We plot 32 channels out of the total 64 channels for P- and S-phases. We observe significant differences among these features, such as localized activations at P- and S-phase arrivals and broad activations during earthquake signals. These features contain characteristic information that is useful to differentiate earthquake signals from background noise and to pick phase arrivals. 

\section{Results}

We evaluate the phase picking performance of EQNet on the previously published and publicly available STanford EArthquake Dataset (STEAD) and evaluate the earthquake detection performance on the 2019 Ridgecrest, CA earthquake sequence. 

\subsection{Benchmark with the STEAD dataset}

\begin{figure}
    \centering
    \begin{subfigure}{0.45\textwidth}
    \includegraphics[width=\textwidth]{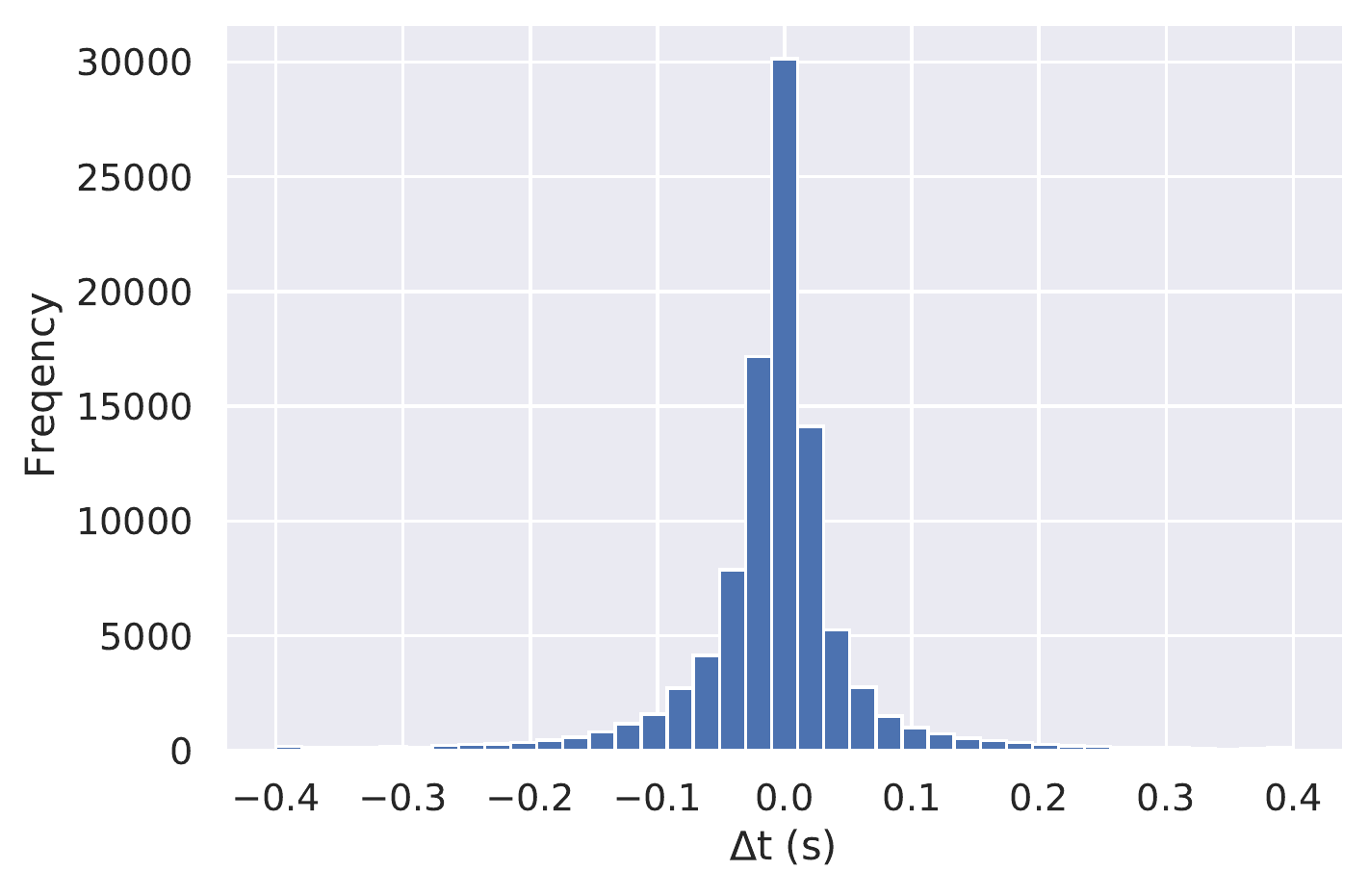}
    \caption{}
    \end{subfigure}
    \begin{subfigure}{0.45\textwidth}
    \includegraphics[width=\textwidth]{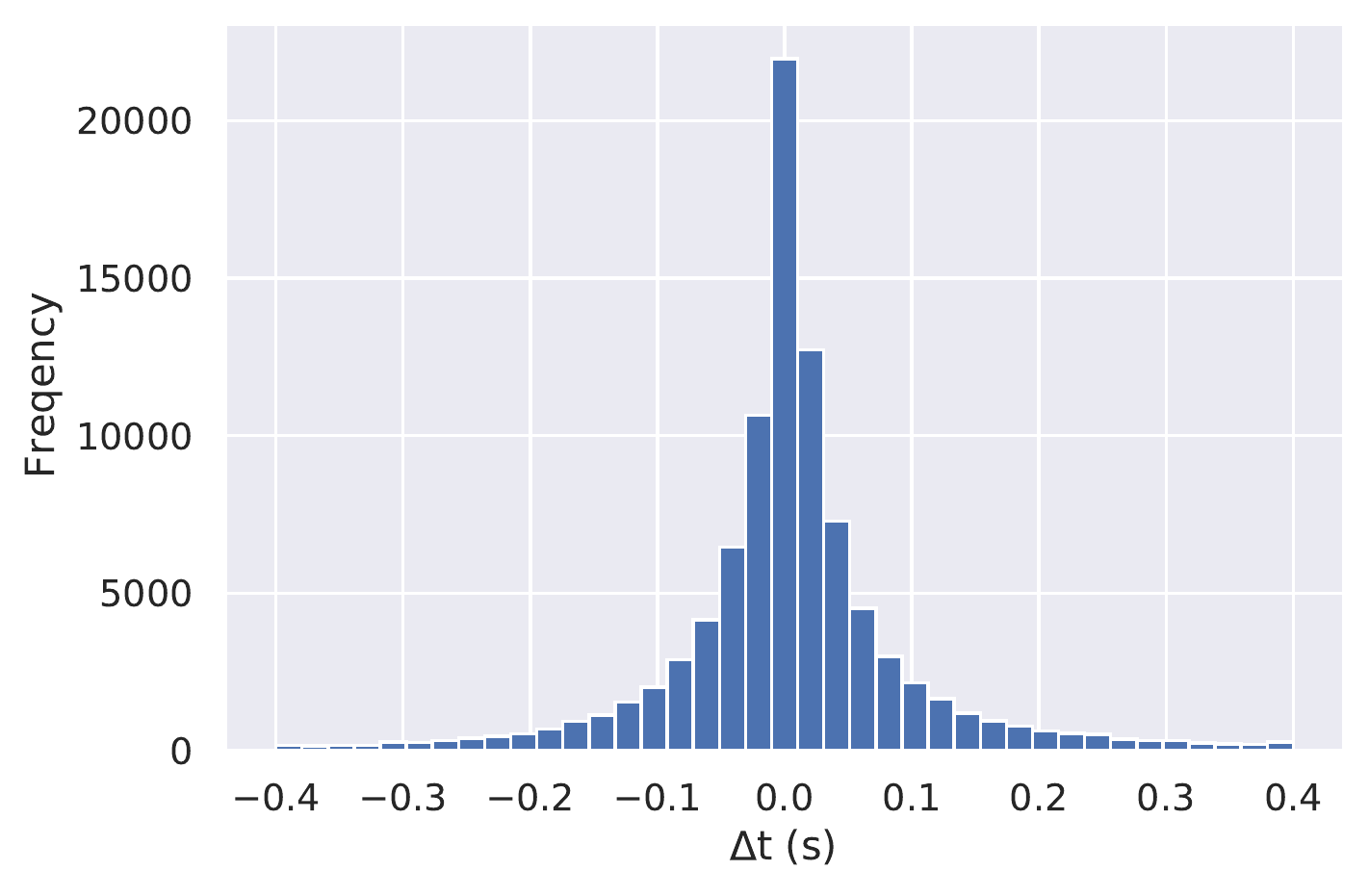}
    \caption{}
    \end{subfigure}
    \caption{Phase picking performance: (a) time residuals of P-picks; (b) time residuals of S-picks.}
    \label{fig:picking_residual}
\end{figure}

STEAD is a global dataset of more than one million seismic waveforms \citep{mousavi2019stanford} with both P- and S-arrival labels that has been used to compare state-of-the-art automatic phase pickers \citep{mousavi2020earthquake}. We selected the same test set ($\sim$120,000 waveforms) used in \citet{mousavi2020earthquake} to evaluate the phase picking performance of EQNet. The waveforms were processed by the feature extraction network and then the phase picking networks to pick P- and S-phase arrivals. We extracted the peaks above a threshold of 0.5 in the predicted activation sequences (\Cref{fig:prediction-sample}(d)) to detect phases and determine arrival times. The distribution of residuals between the predicted arrival times and the manual labels are shown in \Cref{fig:picking_residual}, and the corresponding statistics are listed in \Cref{tab:picking_residual}. The predicted picks that are within 0.5 seconds from the manual labels are counted as true positives. The rest are counted as false positives. Although EQNet uses a simple CNN model, the picking performance approaches the state-of-the-art models PhaseNet \citep{zhu2019phasenet} and EQTransformer \citep{mousavi2020earthquake}. Moveover, although EQNet is trained using data from Northern California only, it generalizes well to this global dataset.

\begin{table}[]
\centering
\caption{Statistics of picking time residuals.}
\label{tab:picking_residual}
\resizebox{0.8\textwidth}{!}{%
\begin{tabular}{lccccccc}
\hline
\multicolumn{2}{l}{}                     & Precision & Recall & F1    & Mean (s) & Std (s) & MAE (s) \\ \hline
\multirow{2}{*}{EQNet}         & P-phase & 0.96      & 0.95   & 0.95  & -0.01    & 0.08    & 0.04    \\
                               & S-phase & 0.92      & 0.92   & 0.92  & 0.01     & 0.11    & 0.07    \\ \hline
\multirow{2}{*}{PhaseNet}      & P-phase & 0.96      & 0.96   & 0.96  & -0.02    & 0.08    & 0.07    \\
                               & S-phase & 0.96      & 0.93   & 0.94  & -0.02    & 0.11    & 0.09    \\ \hline
\multirow{2}{*}{EQTransformer} & P-phase & 0.99      & 0.99   & 0.99  & 0.00     & 0.03    & 0.01    \\
                               & S-phase & 0.99      & 0.96   & 0.98  & 0.00     & 0.11    & 0.01    \\ \hline
\end{tabular}%
}
\end{table}

\subsection{Benchmark on the 2019 Ridgecrest earthquake}

We chose the Ridgecrest earthquake sequence to evaluate EQNet's event detection performance using multi-station waveforms. Ridgecrest is useful for this comparison because in addition to the routine catalog developed by the Southern California Seismic Network (SCSN), \citet{shelly2020high} and \citet{ross2019hierarchical} built two different catalogs using the template matching method, which takes earthquake events in the SCSN catalog as templates and scans the continuous seismic data to detect more small earthquakes. \citet{liu2020rapid} built another catalog using the PhaseNet picker and the REAL \citep{zhang2019rapid} association algorithm. These state-of-the-art catalogs provide good benchmarks for evaluating the performance of our end-to-end method. 

\begin{figure}
    \centering
    \begin{subfigure}{0.45\textwidth}
    \includegraphics[width=\linewidth]{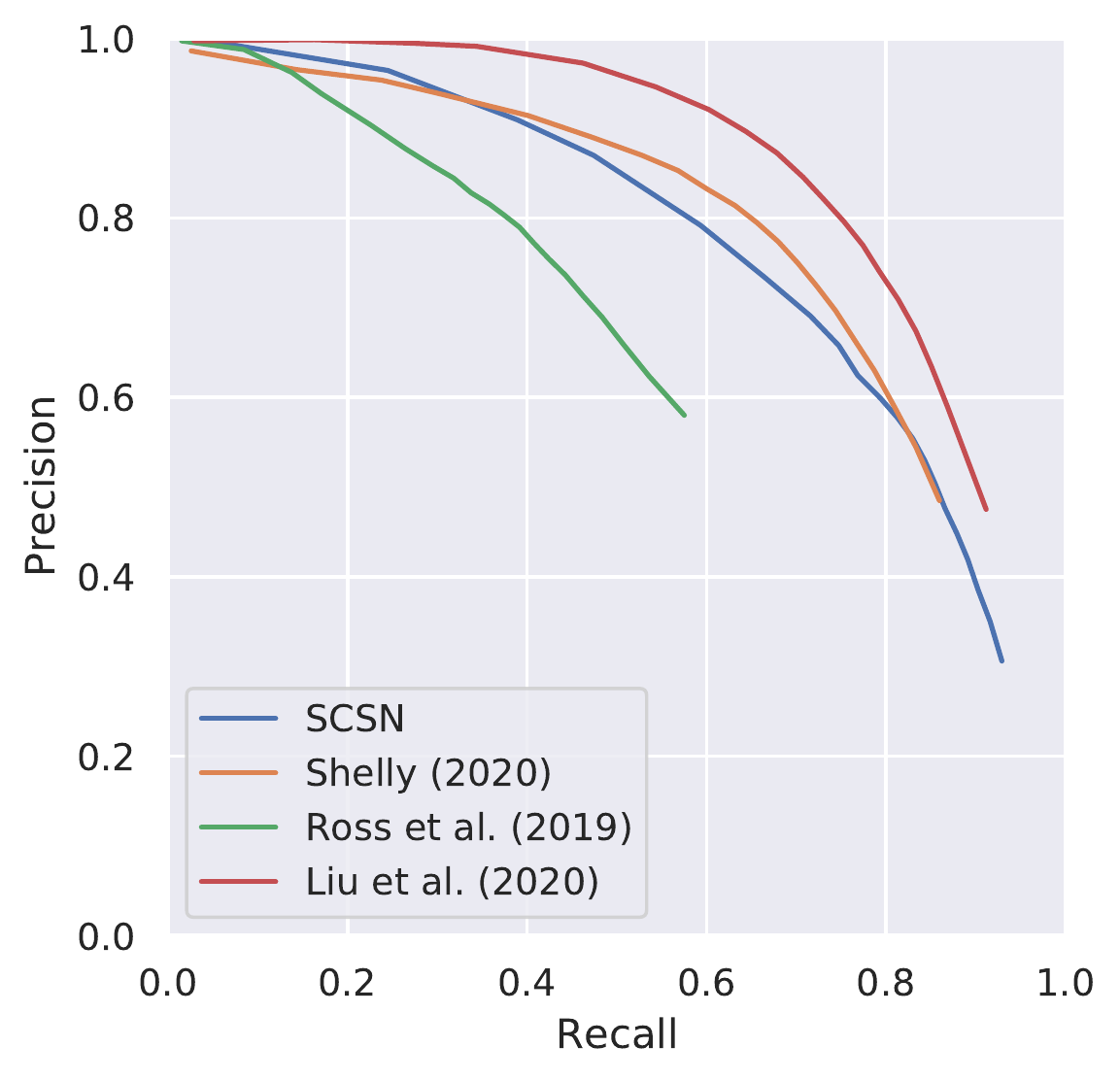}
    \caption{}
    \label{fig:precision-recall}
    \end{subfigure}
    \begin{subfigure}{0.45\textwidth}
    \includegraphics[width=\linewidth]{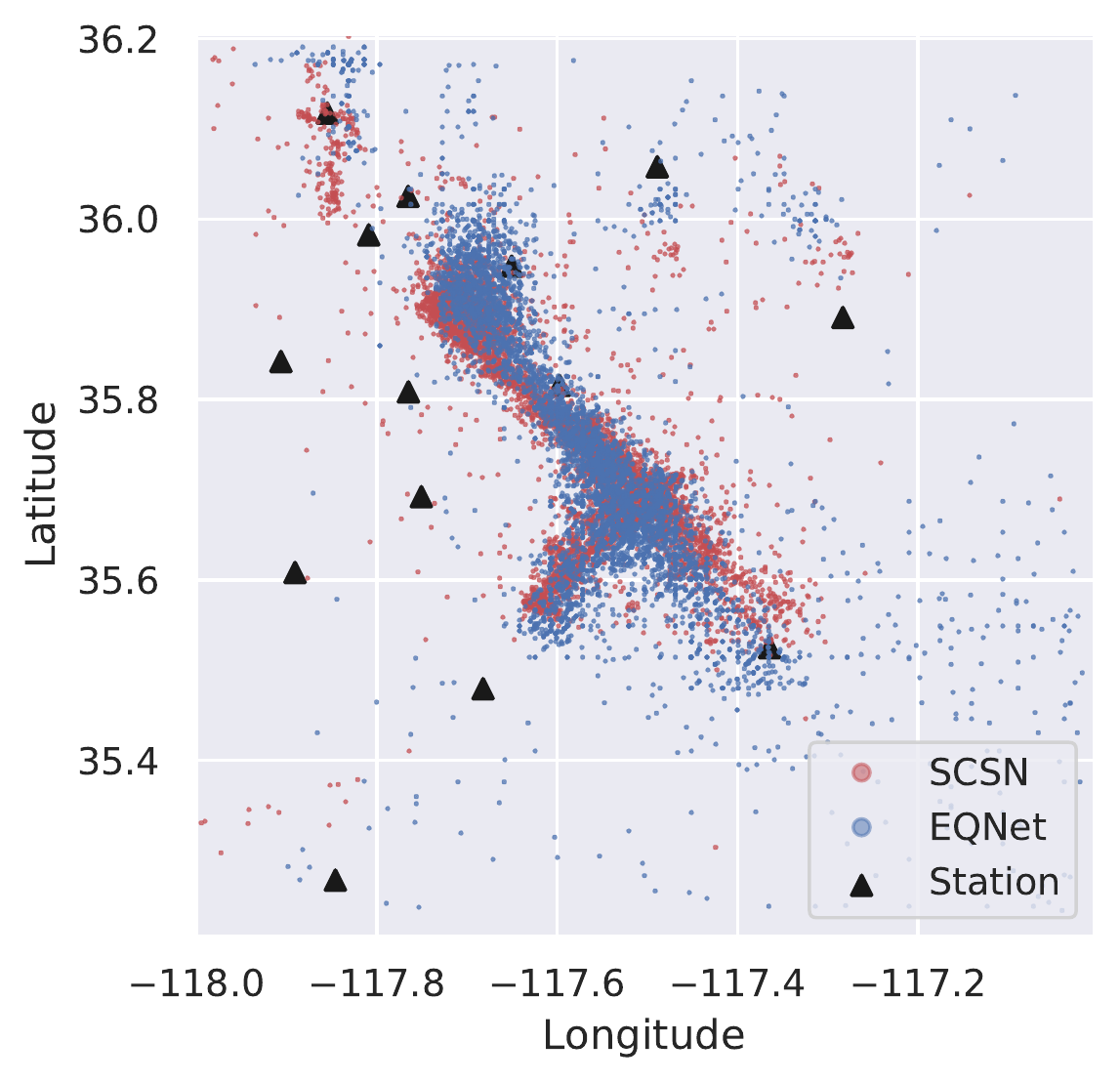}
    \caption{}
    \label{fig:error-loc2d}
    \end{subfigure}
    \begin{subfigure}{0.45\textwidth}
    \includegraphics[width=\linewidth]{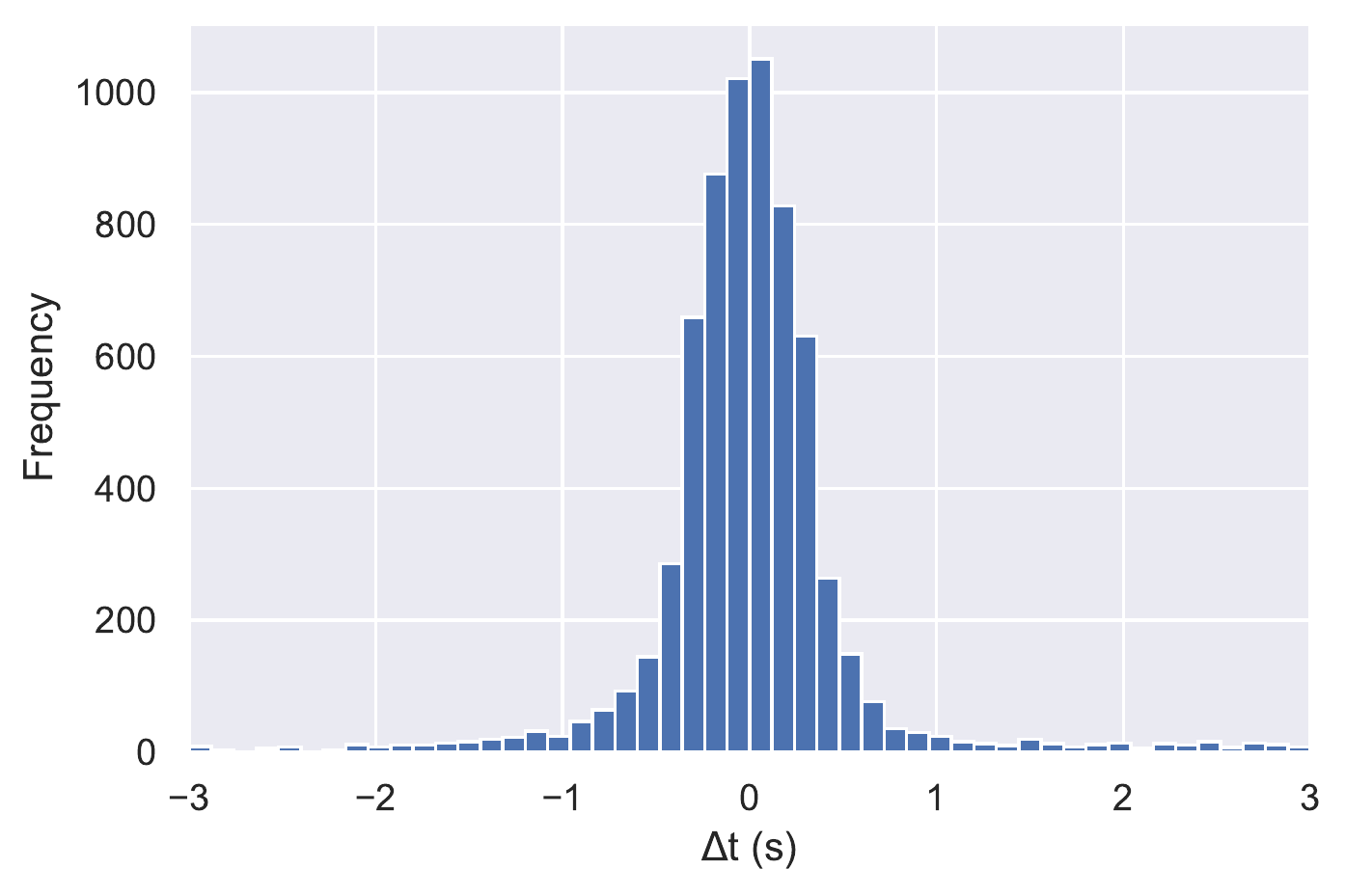}
    \caption{}
    \label{fig:error-time}
    \end{subfigure}
    \begin{subfigure}{0.45\textwidth}
    \includegraphics[width=\linewidth]{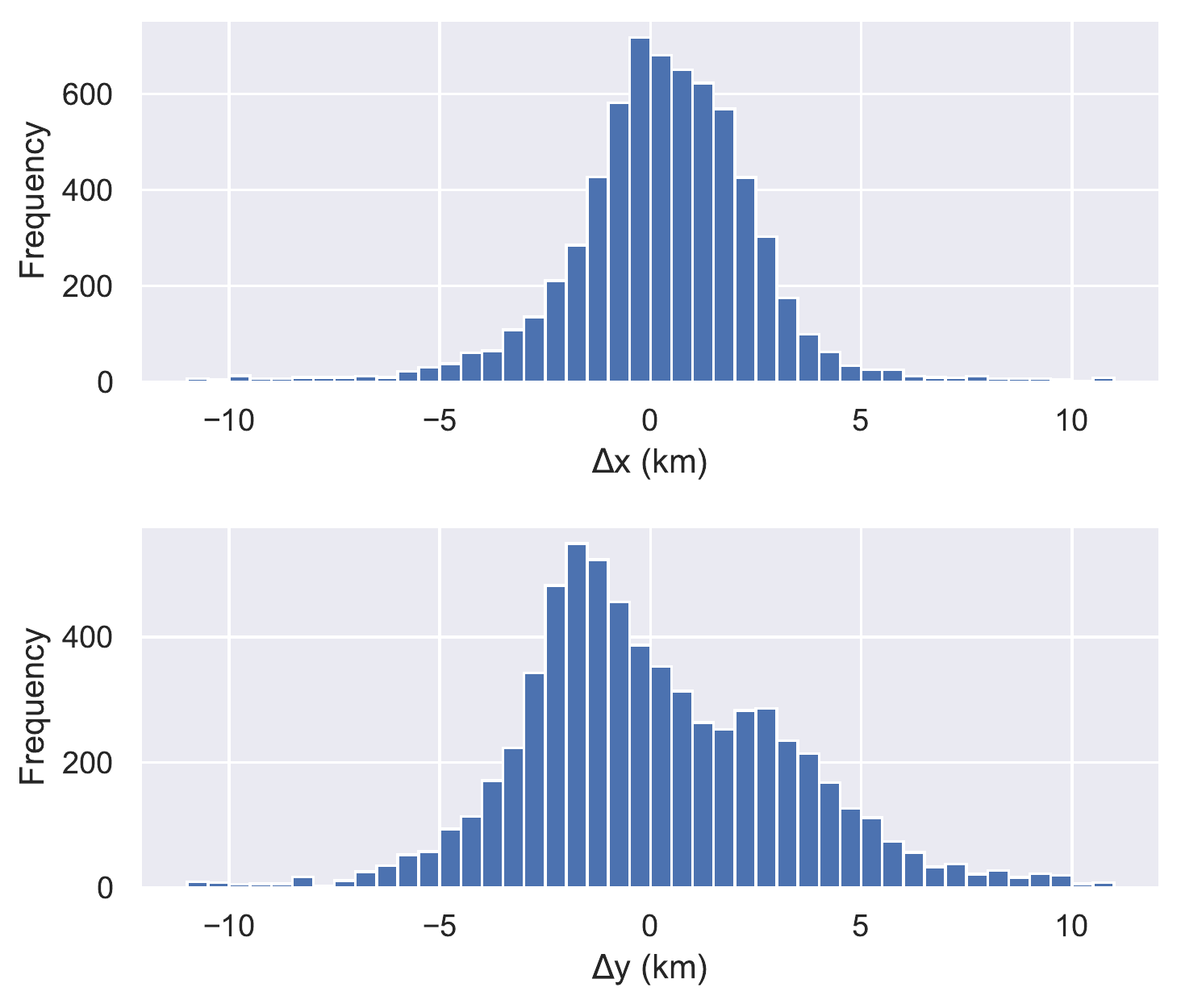}
    \caption{}
    \label{fig:error-loc}
    \end{subfigure}
    \caption{Event detection performance: (a) precision-recall curves calculated by assuming the four state-of-the-art catalogs as ground truth; (b) estimated earthquake locations; (c) error distribution of earthquake time; (d) error distribution of earthquake location compared with that of the SCSN catalog.}
    \label{fig:result}
\end{figure}

We selected five days from July 4th to 8th for the benchmark comparison and downloaded the continuous seismic waveforms from seismic stations within 1 degree of the location (-117.504$W$, 35.705$N$). Earthquakes were frequent and closely spaced in time during these five days, which makes the detection task challenging. During this period, the SCSN catalog reports 8,044 events, \citet{shelly2020high} reports 13,775 events, \citet{ross2019hierarchical} report 23,705 events, and \citet{liu2020rapid} report 12,357 events. We report EQNet's detection number at two activation thresholds: 11,765 events at a threshold of 0.5 and 24,875 events at a threshold of 0.1. Because the ground-truth event number behind these continuous waveforms is unknown, and because each catalog contains false positive and false negative detections due to different algorithms, hyperparameters, and thresholds, it is a challenging task to compare detection performance accurately across catalogs. 
We used a cross-validation test among these catalogs by assuming one catalog as ground truth and calculating the relative performance metrics of the other catalogs. We consider earthquake events whose origin time fall within 3s of the benchmark catalog time as true positives. \Cref{tab:cross-comparison} shows the precision, recall, and F1-score of the cross-validation test. The SCSN catalog (8,044 events) has the highest precision, meaning that most of events in the SCSN catalog also exist in the other catalogs and are more likely to be true earthquakes. The EQNet0.1 catalog (24,875 events) has the highest recall rate meaning that the events in the other catalogs can also be detected by EQNet. The precision and recall values are greatly affected by number of detected earthquakes. The F1-score (the harmonic mean of precision and recall) provides a balanced evaluation criteria of the two measures. \citet{liu2020rapid}'s catalog (12,357 events) and EQNet0.5 catalog (11,765 events) achieves the highest F1-scores. 
Note that this cross-validation result mainly reflects the consistency among catalogs instead of determining which catalog is the most accurate. From the point of view of this paper, the result demonstrates that our end-to-end method achieves a performance similar to that of the other state-of-the-art approaches. Through choosing a range of thresholds between 0.1 - 1, we plot the precision and recall curves of EQNet measured by assuming the other four catalogs as ground truth (\Cref{fig:result}(a)). The curves show good consistency between EQNet's catalog and the catalogs of \citet{liu2020rapid}, \citet{shelly2020high}, and SCSN. The lower similarity to \citet{ross2019hierarchical} is potentially due to either false positive events in that catalog or to false negative events in the EQNet catalog. We further analyze the accuracy of the earthquake origin time and locations of the EQNet0.5 catalog compared with the SCSN catalog (\Cref{fig:result}(b) - (d)). The error distributions of earthquake time and location show that most detected earthquakes have a time error within 1 s and an epicenter error within 6 km. This is not surprising because the event detection network of EQNet only estimates very approximate earthquake locations for detection. We note that these locations are sufficiently accurate to carry out phase association, and to be used as initial location input to conventional location (e.g. hypoinverse \citep{klein2002user}) or relocation (e.g. hypoDD \citep{waldhauser2000double}) algorithms.

\begin{table}[]
\centering
\caption{Cross-validation among four state-of-the-art catalogs and EQNet's catalog}
\label{tab:cross-comparison}
\resizebox{\textwidth}{!}{%
\begin{tabular}{|m{2.7cm}|m{2cm}|C{2.7cm}|C{2.7cm}|C{2.8cm}|C{2.7cm}|C{2.7cm}|C{2.7cm}|}
\hline
\multicolumn{1}{|c|}{} &
   &
  \multicolumn{6}{c|}{Benchmark catalog} \\ \cline{3-8} 
\multicolumn{1}{|c|}{\multirow{-2}{*}{Testing catalog}} &
  \multirow{-2}{*}{Coefficient} &
  SCSN &
  Shelly (2020) &
  Ross et al. (2019) &
  Liu et al. (2020) &
  EQNet0.5 &
  EQNet0.1 \\ \hline
\cellcolor[HTML]{EFEFEF} &
  Precision &
  \cellcolor[HTML]{EFEFEF}1.0 &
  \textbf{0.84} &
  \cellcolor[HTML]{EFEFEF}\textbf{0.90} &
  \textbf{0.83} &
  \cellcolor[HTML]{EFEFEF}\textbf{0.83} &
  \textbf{0.93} \\
\cellcolor[HTML]{EFEFEF} &
  Recall &
  \cellcolor[HTML]{EFEFEF}1.0 &
  0.49 &
  \cellcolor[HTML]{EFEFEF}0.30 &
  0.53 &
  \cellcolor[HTML]{EFEFEF}0.55 &
  0.31 \\
\multirow{-3}{*}{\cellcolor[HTML]{EFEFEF}\begin{tabular}[c]{@{}l@{}}SCSN\\ (8,044)\end{tabular}} &
  F1 score &
  \cellcolor[HTML]{EFEFEF}1.0 &
  0.62 &
  \cellcolor[HTML]{EFEFEF}0.45 &
  0.64 &
  \cellcolor[HTML]{EFEFEF}0.67 &
  0.46 \\ \hline
\cellcolor[HTML]{EFEFEF} &
  Precision &
  \cellcolor[HTML]{EFEFEF}0.49 &
  1.0 &
  \cellcolor[HTML]{EFEFEF}0.79 &
  0.72 &
  \cellcolor[HTML]{EFEFEF}0.68 &
  0.86 \\
\cellcolor[HTML]{EFEFEF} &
  Recall &
  \cellcolor[HTML]{EFEFEF}0.84 &
  1.0 &
  \cellcolor[HTML]{EFEFEF}0.45 &
  0.77 &
  \cellcolor[HTML]{EFEFEF}0.77 &
  0.49 \\
\multirow{-3}{*}{\cellcolor[HTML]{EFEFEF}\begin{tabular}[c]{@{}l@{}}Shelly (2020)\\ (13,775)\end{tabular}} &
  F1 score &
  \cellcolor[HTML]{EFEFEF}0.62 &
  1.0 &
  \cellcolor[HTML]{EFEFEF}0.57 &
  0.75 &
  \cellcolor[HTML]{EFEFEF}0.72 &
  0.62 \\ \hline
\cellcolor[HTML]{EFEFEF} &
  Precision &
  \cellcolor[HTML]{EFEFEF}0.30 &
  0.45 &
  \cellcolor[HTML]{EFEFEF}1.0 &
  0.42 &
  \cellcolor[HTML]{EFEFEF}0.39 &
  0.58 \\
\cellcolor[HTML]{EFEFEF} &
  Recall &
  \cellcolor[HTML]{EFEFEF}0.90 &
  0.79 &
  \cellcolor[HTML]{EFEFEF}1.0 &
  0.80 &
  \cellcolor[HTML]{EFEFEF}0.79 &
  \textbf{0.58} \\
\multirow{-3}{*}{\cellcolor[HTML]{EFEFEF}\begin{tabular}[c]{@{}l@{}}Ross et al. (2019)\\ (23,705)\end{tabular}} &
  F1 score &
  \cellcolor[HTML]{EFEFEF}0.45 &
  0.57 &
  \cellcolor[HTML]{EFEFEF}1.0 &
  0.55 &
  \cellcolor[HTML]{EFEFEF}0.52 &
  0.58 \\ \hline
\cellcolor[HTML]{EFEFEF} &
  Precision &
  \cellcolor[HTML]{EFEFEF}0.53 &
  0.77 &
  \cellcolor[HTML]{EFEFEF}0.80 &
  1.0 &
  \cellcolor[HTML]{EFEFEF}0.80 &
  0.91 \\
\cellcolor[HTML]{EFEFEF} &
  Recall &
  \cellcolor[HTML]{EFEFEF}0.83 &
  0.72 &
  \cellcolor[HTML]{EFEFEF}0.42 &
  1.0 &
  \cellcolor[HTML]{EFEFEF}0.75 &
  0.48 \\
\multirow{-3}{*}{\cellcolor[HTML]{EFEFEF}\begin{tabular}[c]{@{}l@{}}Liu et al. (2020)\\ (12,357)\end{tabular}} &
  F1 score &
  \cellcolor[HTML]{EFEFEF}0.64 &
  \textbf{0.75} &
  \cellcolor[HTML]{EFEFEF}0.55 &
  1.0 &
  \cellcolor[HTML]{EFEFEF}\textbf{0.77} &
  \textbf{0.63} \\ \hline
\cellcolor[HTML]{EFEFEF} &
  Precision &
  \cellcolor[HTML]{EFEFEF}\textbf{0.55} &
  0.77 &
  \cellcolor[HTML]{EFEFEF}0.79 &
  0.80 &
  \cellcolor[HTML]{EFEFEF}1.0 &
  1.0 \\
\cellcolor[HTML]{EFEFEF} &
  Recall &
  \cellcolor[HTML]{EFEFEF}0.83 &
  0.68 &
  \cellcolor[HTML]{EFEFEF}0.39 &
  0.75 &
  \cellcolor[HTML]{EFEFEF}1.0 &
  0.47 \\
\multirow{-3}{*}{\cellcolor[HTML]{EFEFEF}\begin{tabular}[c]{@{}l@{}}EQNet0.5\\ (threshold=0.5)\\ (11,765)\end{tabular}} &
  F1 score &
  \cellcolor[HTML]{EFEFEF}\textbf{0.66} &
  0.72 &
  \cellcolor[HTML]{EFEFEF}0.52 &
  \textbf{0.77} &
  \cellcolor[HTML]{EFEFEF}1.0 &
  0.64 \\ \hline
\cellcolor[HTML]{EFEFEF} &
  Precision &
  \cellcolor[HTML]{EFEFEF}0.30 &
  0.48 &
  \cellcolor[HTML]{EFEFEF}0.58 &
  0.47 &
  \cellcolor[HTML]{EFEFEF}0.47 &
  1.0 \\
\cellcolor[HTML]{EFEFEF} &
  Recall &
  \cellcolor[HTML]{EFEFEF}\textbf{0.93} &
  \textbf{0.86} &
  \cellcolor[HTML]{EFEFEF}\textbf{0.58} &
  \textbf{0.91} &
  \cellcolor[HTML]{EFEFEF}1.0 &
  1.0 \\
\multirow{-3}{*}{\cellcolor[HTML]{EFEFEF}\begin{tabular}[c]{@{}l@{}}EQNet0.1\\ (threshold=0.1)\\ (24,875)\end{tabular}} &
  F1 score &
  \cellcolor[HTML]{EFEFEF}0.46 &
  0.62 &
  \cellcolor[HTML]{EFEFEF}\textbf{0.58} &
  0.62 &
  \cellcolor[HTML]{EFEFEF}0.64 &
  1.0 \\ \hline
\end{tabular}%
}
\end{table}

\section{Discussion} 

Multi-stage earthquake detection workflows consisting of a sequence of processing tasks, such as, phase detection/picking, association, location, and characterization, have proven to be an effective approach for earthquake monitoring; however, each step of this multi-stage approach involves information loss, such as weak signals undetected because they fall below a threshold at the phase picking step. An end-to-end approach without intermediate steps avoids this information loss by using seismic waveforms to improve overall earthquake detection performance. We have developed an end-to-end neural network architecture for earthquake detection to solve the phase picking and association tasks simultaneously. The end-to-end architecture consists of three neural networks for feature extraction, phase picking and event detection. These networks are jointly optimized during training to maximize their performance. The end-to-end training approach eliminates separate hyper-parameter tuning for each step in conventional earthquake detection workflows and avoids information loss in each step. EQNet combines the advantages of effective feature extraction of deep neural networks with the high detection robustness of array-based methods by processing waveforms from stations across a seismic network. The good performance of both the phase picker and the event detector networks demonstrates that the extracted features (\Cref{fig:prediction-sample}(b) and (c)) contain characteristic information of seismic waveforms. The event detection network combines these features to recognize coherent patterns across multiple stations and thus improve earthquake detection, particularly for the important case of weak signals.

In contrast to models that are trained on a fixed set of stations and neglect station location information \citep{zhang2020locating, yang2021simultaneous}, we design a shift-and-stack module that serves a similar function as the back-projection method commonly employed in earthquake detection and imaging \citep{kao2004source, inbal2015imaging, ben2015basic}. This module incorporates physical constraints of travel times determined by station locations and wavespeed models, making EQNet generalizable to other research regions using different station geometries and wavespeed models. In this work, for example, we demonstrated that the model trained with data from Northern California generalizes well on the Ridgecrest earthquake sequence in Southern California. The sampling module enables generalizability at some computational cost because the event detection network needs to process a collection of shifted features sampled at different spatial locations. To reduce the computational effort, we engineer a deep backbone network, which processes the input waveforms only once to extract features, and a shallow event detection network, which efficiently processes a group of shifted features in parallel. Conventional back-projection methods for local earthquake detection and location have strict requirements of coherent waveforms and polarity, accurate velocity model, and can come at high computational cost from dense grid sampling \citep{beskardes2018comparison}. Filtered waveforms and characteristic functions, such as the complex envelope \citep{gharti2010automated}, STA/LTA \citep{fischer2020microseismic}, and kurtosis \citep{langet2014continuous}, are used to improve stacking results and to reduce computational cost. The features extracted by EQNet provide a rich representation of characteristic functions that are informed by data and are optimized for the relevant learning tasks (\Cref{fig:prediction-sample}(b) and (c)), thus EQNet achieves good performance without either strict requirements or preprocessing. In the experiments above, we applied EQNet directly to the raw waveforms, using an assumed uniform velocity model of 6 km/s for P waves and 3.4 km/s for S waves.  We applied grid-search to latitude and longitude components with a spatial interval of ~4 km. Beyond this proof-of-concept study, it will be possible to use more realistic velocity models and/or improved travel-time predictions to improve the output. The computational time required to process 24 hours of continuous data from 42 three-component seismometers is around 7 minutes on a Tesla V100-SXM2-32GB-LS GPU, and because the problem is embarrassingly parallel, the processing can be easily distributed over multiple GPUs. 

Earthquake monitoring bears some similarity to object detection in computer vision, which aims to locate and classify objects in an image \citep{girshick2014rich, girshick2015fast, ren2015faster, liu2016ssd, redmon2016you}. CNN-based object detection methods, such as YOLO \citep{redmon2016you}, scan a set of coarse grids to classify object categories and predict bounding boxes. Similarly, EQNet detects earthquakes by scanning a spatial-temporal space. Because most earthquakes, and particularly small earthquakes, can be approximated as a point source, we do not need to predict bounding boxes of different shapes as is the case in object detection. Our end-to-end approach opens a new pathway for the earthquake detection problem at the seismic network level. It resembles the way analysts monitor earthquakes by examining a collection of waveforms recorded across multiple stations. New algorithms and architectures in the rapidly evolving field of object detection will likely improve this end-to-end approach for earthquake monitoring in the near future.

\section{Conclusions}

Deep learning is an effective approach for earthquake detection; however, currently a dominant approach in deep learning models for seismic monitoring is to process seismic waveforms one station at a time, which can not realize the full potential of network-based monitoring. To do that, deep learning needs to be combined with effective association methods to aggregate information from multiple stations in a seismic network to detect earthquakes and reduce false positives. We have developed an end-to-end model, EQNet, that combines detection and association for array-based earthquake detection. EQNet consists of three neural networks for feature extraction, phase picking, and event detection, and jointly optimizes these tasks to improve overall detection performance. EQNet processes multiple waveforms from a seismic network to improve detection sensitivity for small earthquakes. Application to the STEAD dataset demonstrates that EQNet can pick P- and S-phases as effectively as other state-of-the-art deep learning models. The cross-validation result on the 2019 Ridgecrest earthquake sequence demonstrates that EQNet achieves good earthquake detection performance and generates earthquake catalogs consistent with four state-of-the-art approaches. EQNet represents a new strategy for developing array-based deep learning models to improve earthquake detection. 

\section*{Acknowledgements}
Waveform data, metadata, or data products for this study were accessed through the Northern California Earthquake Data Center (NCEDC) and the Southern California Earthquake Data Center (SCEDC). The EQNet catalogs are available in Open Science Framework (\url{https://doi.org/ 10.17605/OSF.IO/HCVJX}). This work was supported by the Department of Energy (Basic Energy Sciences; Award DE-SC0020445).

\clearpage
\bibliographystyle{apacite}
\bibliography{reference.bib}

\end{document}